\documentclass{my_aa}
\usepackage{natbib}
\usepackage{epsfig}
\usepackage{txfonts}

\def\ppmm{$\pm$}
\begin{document}
\title{Hubble Space Telescope time-series photometry of the planetary transit of HD189733: no moon, no rings, starspots}
\author{Fr\'ed\'eric Pont$^{1}$, Ronald L. Gilliland$^{2}$, Claire Moutou$^{3}$, David Charbonneau$^{4}$, Fran\c cois Bouchy$^{5}$, Timothy M. Brown$^{6}$,  Michel Mayor$^{1}$, Didier Queloz$^{1}$, Nuno Santos$^{7}$, Stephane Udry$^{1}$}
\offprints{frederic.pont@obs.unige.ch}
\institute{$^{1}$Geneva University Observatory, 1290-Sauverny, Switzerland\\
$^{2}$Space Telescope Science Institute,3800 San Martin Drive, Baltimore, MD 21128, USA\\
$^{3}$Laboratoire d'Astrophysique de Marseille, Traverse du Siphon, 13376 Marseille Cedex 12, France \\
$^{4}$Harvard-Smithsonian Center for Astrophysics, 60 Garden Street, Cambridge, MA 02138, USA\\
$^{5}$Laboratoire d'Astrophysique de Paris, 98bis Bd Arago, 75014 Paris, France\\
$^{6}$Las Cumbres Observatory, Goleta CA 93117, USA\\
$^{7}$Centro de Astrof\' \i sica da Universidade do Porto, Rua das Estrelas, 4150-762 Porto, Portugal\\}  

\date{Received date / accepted date}

   \authorrunning{F. Pont et al.}
   \titlerunning{Hubble Space Telescope times-series photometry of HD189733}
\abstract{We monitored three transits of the giant gas planet around the nearby K dwarf HD 189733 with the ACS camera on the Hubble Space Telescope. The resulting very-high accuracy lightcurve (signal-to-noise ratio near 15000 on individual measurements, 35000 on 10-minute averages) allows a direct geometric measurement of the orbital inclination, radius ratio and scale of the system: $i = 85.68 \pm 0.04$, $R_{pl}/R_*=0.1572 \pm 0.0004$, $a/R_*=8.92 \pm 0.09$. We derive improved values for the stellar and planetary radius, $R_*=0.755 \pm 0.011$\ R$_\odot$, $R_{pl}=1.154\pm 0.017$\ R$_J$, and the transit ephemerides, $T_{tr}=2453931.12048 \pm 0.00002 + n \cdot 2.218581 \pm 0.000002$. The HST data also reveal clear evidence of the planet occulting spots on the surface of the star. At least one large spot complex ($>$ 80000 km) is required to explain the observed flux residuals and their colour evolution. This feature is compatible in amplitude and phase with the variability observed simultaneously from the ground. No evidence for satellites or rings around HD 189733$b$ is seen in the HST lightcurve. This allows us to exlude with a high probability the presence of Earth-sized moons and Saturn-type debris rings around this planet. The timing of the three transits sampled is stable to the level of a few seconds, excluding a massive second planet in outer 2:1 resonance.

\keywords{planetary systems -- stars: individual: HD 189733 -- techniques: photometric}}

\maketitle

\section{Introduction}

%
%

The star HD 189733 is transited by a close planetary companion (Bouchy et al. 2005). The brightness of the host star ($V$=7.7 mag), the short period (2.2 days) and the large planet-to-star radius ratio ($R_{pl}/R_* \simeq$0.15)  make it the most favourable system known for detailed studies. In the year following the discovery of its transiting planet, HD 189733 has already been subjected to many follow-up observations, including extensive ground-based transit photometry (Bakos et al. 2006, Winn et al. 2007), measurement of the spin-orbit angle using the Rossiter-McLaughlin effect (Winn et al. 2006), measurement and mapping of the infrared flux distribution of the planet with the secondary eclipse (Deming et al. 2006, Grillmair et al. 2007, Knutson et al. 2007).

A precise knowledge of the radius of the planet is essential to all applications. From the transit lightcurve, one can measure the geometry of the system, hence the orbital inclination, the size of the star
 and the size of the planet. However, to first order the orbital inclination and the stellar radius can compensate almost perfectly. With ground-based observations, it is extremely difficult to lift completely the degeneracy between orbital inclination and stellar radius, because of systematics induced by the Earth's atmosphere.  The spectacular precision reached by space-based transit time series was illustrated with the HST/STIS lightcurve of HD 209458 by Brown et al. (2001).

With an accuracy in the $10^{-4} - 10^{-5}$ range, second-order features in the transit lightcurve can be detected, such as the transit of a satellite of the planet, or the presence of Saturn-type debris rings.

%
%

\section{Observations}

We have observed HD 189733 with the Hubble Space Telescope during 3 visits of 5 orbits each, using the Advanced Camera for Surveys (ACS) in HRC mode with the grism G800L (program GO-10923). In this configuration, the first-order spectrum ranges from 5500 \AA\ to 10500 \AA\ with $\sim 40$ \AA\ per pixel. The first visit occured on May 22, 2006 (JD 245877.718 to 878.092), the second on May 26 (882.115 -- 882.408) and the third on July 14 (930.946 -- 931.240). 

To achieve the highest possible signal-to-noise ratio (S/N), we observed with relatively long exposures (25s), allowing the ACS detector to saturate by a factor about 5 at the wavelengths near 700 nm where it is most sensitive. Prior experience doing time-series photometry with the ACS showed that with proper analysis, Poisson-limited photometry can be obtained even with images that are much more saturated than this. This is done by integrating over enough pixels to capture essentially all electrons generated by the stellar photons, even though they may have diffused away from the pixel where they were created. For crowded  field photometry, this approach can be problematic because of leakage from adjacent star images, but for spectroscopy on the bright HD189733 this is not a problem.

We obtained extremely high S/N spectrophotometry in a time-series mode with a cadence of about 60s. Since the transit duration is more than the length of an HST orbit and the target star can be seen for less than half of each orbit, three visits were required to obtain full phase coverage of the transit. We accumulated a total of 675 exposures for the three visits.
 Each visit contains two full HST orbits before the transit, two that are all or partly inside the transit, and one that falls entirely after the transit. Observations on both sides of the transits are necessary for each visit so that clear out-of-transit baselines can be established. Also, the  first HST orbit often exhibits unique systematics as the telescope and instrument stabilize at the new pointing. The first orbit of each visit is not utilized in the  final time series analyses.

\section{Reductions}

\label{reduc}

Our analysis started with the images provided by the STScI pipeline. 
These are corrected with bias subtraction,
dark current subtraction after scaling to individual
exposure times, and converted from digital units to electrons scaling by
the detector gain.  Retrievals from the STSci archive were made
only after final biases and darks were available.

These images are not flat-fielded for grism data.  
For the very similar GO/DD-10441 observations of TrES-1 (Brown et al., in preparation)
in which HRC grism spectra at high S/N were obtained at nearly
the same place on the detector, but without the complication
of saturation and bleeding, it was found that use of a wavelength-dependent 
flat field provided no benefit.
Facing the extra complication here of saturated data and
needing to take into account the flat fields as a function
only of electron origination, not current location, we simply
have not applied flat fields for these data.
  For our purposes of deriving differential time series
over time, in which all of the spectra have been taken at
nominally the same pointing, and for which the extractions
are averaged over a large domain (over 5000 pixels for the global first
order spectrum extraction), we would not expect
much sensitivity to flat fielding errors.  If there were
no motion of the spectra relative to the detector as a
result of HST guiding errors, there should be no need to
flat field.  These guiding errors, while not zero, are
quite small ($\sim1/10$ pixel), and at a later step we decorrelate with
vectors that remove any small noise resulting from lack
of flat fielding.

\subsection{Extraction}

The images consist of a direct image, a first-order spectrum, and
a partial second-order spectrum, inclined diagonally across the detector.
The signal of saturated pixels is spread along columns (i.e. in the $y$ direction).
  Analyses start with measurement of $x,y$ position for the
single direct image taken at the start of each visit.
The calibration given in ACS-ISR 03-07 is used to provide
a wavelength for each position along the spectrum trace.
  The global first-order spectrum extraction involves
a sum over 173 columns that span wavelengths of about
5370 to 10690 \AA.  The extractions  involve sums over
columns, rather than attempting extractions in the
pure cross-dispersion direction.  For non-saturated
data this results in a moderate loss of spectral
resolution, for saturated data the resolution loss is
inherent in the data anyhow.  We define an arbitrary
data value at which to extend the summations along
columns, stopping the summation the first time the
data drop below this value.  A number of trials are
run adopting a range of these arbitrary sum limiting
values, and the one providing highest S/N (after
applying all the corrections discussed later) is
adopted.  For HD 189733 a stopping value of 6,000
electrons works best (peak
intensities would have been over $10^6$ in
the absence of saturation, thus the sums extend to
levels of less than 1\% relative intensity.)
  A first extraction is done using the above
procedure, and intensities are normalized by the
mean for all points (except first) in orbits
two and five -- this provides an initial light curve.
  We
perform a cosmic ray elimination step by
sigma clipping data values outside the saturated
data region (with a buffer of two pixels), over the
full stack of images in each visit -- after normalizing
out the intensity variations during transit via use of
the preliminary light curve.  After replacing data
values affected by cosmic rays the normalization is
reset.  The sigma clipping is done after providing
a model for the intensity variations based on a linear
regression with the six external variables to be
described below in the decorrelation section.
Cosmic ray
elimination is not very important for this project
that used short exposure times and accumulated over
$5\times 10^8$ source electrons each exposure.
  The latter procedure could only catch cosmic rays
outside the saturated pixel domain.  To search for
cosmic rays within the saturated pixel region we formed
stacks of column-summed intensities, and performed the
same data modelling and sigma clipping step as above.
This resulted in trivial improvements for 2 of 3 visits,
and a minor loss for the third.  We have kept this step
although it neither helped nor harmed the overall
result to any significant extent. 

For a specific application (see Section~\ref{spot}), we also extracted sub-sections
of the spectrum in 500~\AA\ wavelength intervals, restricting the extraction to specific
ranges in $y$.

\subsection{Decorrelation}

  It is important to establish by how much the
spectrum moves relative to the detector exposure-to-exposure, 
and at the same time establish records of
other characteristics such as measures of the
cross-dispersion width and the rotation of the
spectrum relative to the detector.  Such external
vectors become candidates for performing decorrelations
with the intensity vectors.

  For each time step a one-dimensional Gaussian was
fit to all columns in the first- and second-order
spectra for which (a) the core does not saturate,
(b) the column is not next to one that saturates,
and (c) the column intensity sum reaches $10^5$ electrons.
The results of these fits are then averaged together.
After subtracting the mean this provides time series
records of: (1) mean $y$-position of the spectra,
(2) a measure of the spectrum width -- Gaussian sigma,
and (3) a rotation formed by differencing mean $y$-position
in first and last 10 columns, which for these data are
separated by about 65 columns.
  A measure of the spectrum shifts in $x$ are derived
by forming a one-dimensional vector of column-summed
intensities over $x$, averaged over all the exposures
in a visit.  Pixel-by-pixel in $x$ the intensity derivative
in $x$ is formed, and note is taken of the region (columns
2-15 in practice) that has both decent signal and strong
derivatives.  Then averaged over these 14 columns the
relative intensities in individual spectra in combination
with the $dI/dx$ derivative provide delta-$x$ shifts for
each exposure.
The spectrum shifts in $x$ with a characteristic
pattern each orbit by about 0.1 pixel peak to peak.
A somewhat larger offset occurs in $y$, with both
a long-term drift and within orbit variations present.
The cross-dispersion width becomes more variable with
each successive orbit.  Rotation maintains a characteristic
pattern each orbit. 

  Decorrelations are performed by evaluating a
multi-linear regression on these vectors simultaneously:
$x$-position, $y$-position, width, rotation, HST orbit phase,
and overall time.  No attempt is made to account for
the fact that several of these external vectors show
mutual correlations.  The regression is based on
a least-squares fit using all of the points outside
of transit, ignoring orbit~1.  Then the
full set of data points is divided by the fit to
provide corrections for noise that correlate with
these external vectors.  

The decorrelations generally remove noise with characteristic
time scales smaller than the HST orbit, plus an overall linear trend.
For the average number of 130 data points used to establish the decorrelation fits
in each visit,
the time series standard deviation improves from a geometric mean over
the three visits of $277 \cdot 10^{-6}$ before, to $67 \cdot 10^{-6}$ after decorrelation.
The mean absolute value of individual linear correlation coefficients are:
x-position (0.22), y-position (0.48), width (0.35), rotation (0.53),
orbit phase (0.32), and overall time (0.84).

No additional sigma-clipping of the data has been
performed. Exposure 65 of visit 02 is affected by a strong
cosmic ray and was not used.

\subsection{Systematics after decorrelation}

\label{syscor}

Since a primary objective of this program is to measure deviations from the transit shape
at the level of a fraction of a millimagnitude, it is crucial to
estimate the level of systematics that could remain in the data
after decorrelation. One method is to use decorrelation trials in which some out-of-transit orbits are
not used in the regression to establish the decorrelation.
Here there are two visits, the first (V1) and second (V2) that have two
orbits either after (V1) or before (V2) the transit.
The exercise is to step through not using either orbits
2 or 3 of V2, or orbits 4 or 5 of V1.  The residuals
are then compared to those in which all of the data
were used.  The results were very clear.  In the four
orbits (total) not used in decorrelations the largest
deviations remained under $10^{-4}$, with means typically
an order of magnitude smaller still.  In no case did
any residual show deviations as large as
those seen multiple times for orbits inside the transits. This implies that the variations are real and not due to instrumental systematics.

A second method, this time to check if some real signal could have been removed by 
the decorrelation, is to insert features in the data before
decorrelation, and check how they are recovered.
Decorrelation trials in which signals similar to those
in V2 (rise from 0.0 to +0.5 mmag in about 10 points
followed by a drop back to 0.0 in 10 points) are added
to the time series before decorrelation.  We chose to
add these to either orbits 2 or 3 of V2, or 4 or 5
of V1 (one at a time).  In 4 of 6 total cases this
triangle-shaped perturbation was added to centers of
orbits, in the other 2 it was added near the end of
orbits.  In each case the signal is preserved through
the decorrelation step at the 80-90\% level.  It is
not surprising that there is a slight reduction in
height resulting from having added in a positive
signal.

The above two tests show that:  (a) the out-of-transit
orbits stay less noisy, and show no systematics even
when not used to establish the decorrelation, and (b)
if real systematic deviations had been present that
last only a fraction ($\sim$1/3rd) of an orbit, these would
only be affected at the 10-20\% level by decorrelation.
This adds significantly to believing that the residual features
seen during the in-transit HST orbits are real, and
likely reflect flux inhomogeneities on the surface of the star.

We also tested that our results are not
sensitive to the size of the extraction box used, by repeating the 
analysis with a limit at 2,000 electrons (instead of 6,000).

\subsection{Absolute flux}

\label{abs}

Previous experience with ACS and HST show that the variability
of ACS over a few weeks is about 0.003 to 0.004 in fractional flux calibration. Much of the deviation
is correlated with known variation of focus. After correcting for these variations, 
an epoch-to-epoch precision of $\sim < 0.002$ can be obtained.

  The visit-to-visit changes of total intensity are,
based on the mean of orbits 2 and 5 without the first point in orbits:
+0.005 for visit V2 relative to visit V1, and +0.007 for V3 relative to V1.  
The Poisson limit on these would be less than 5$\cdot 10^{-6}$.
There is a caveat that needs to apply to V3 -- this
is after the switch to Side 2 electronics in early July 2006.  There
could be a minor change in operating temperature.
We don't expect a change, but  there is a potential zero-point 
difference for visit V3.

%


%

%


\begin{figure*}
\resizebox{14cm}{!}{\includegraphics{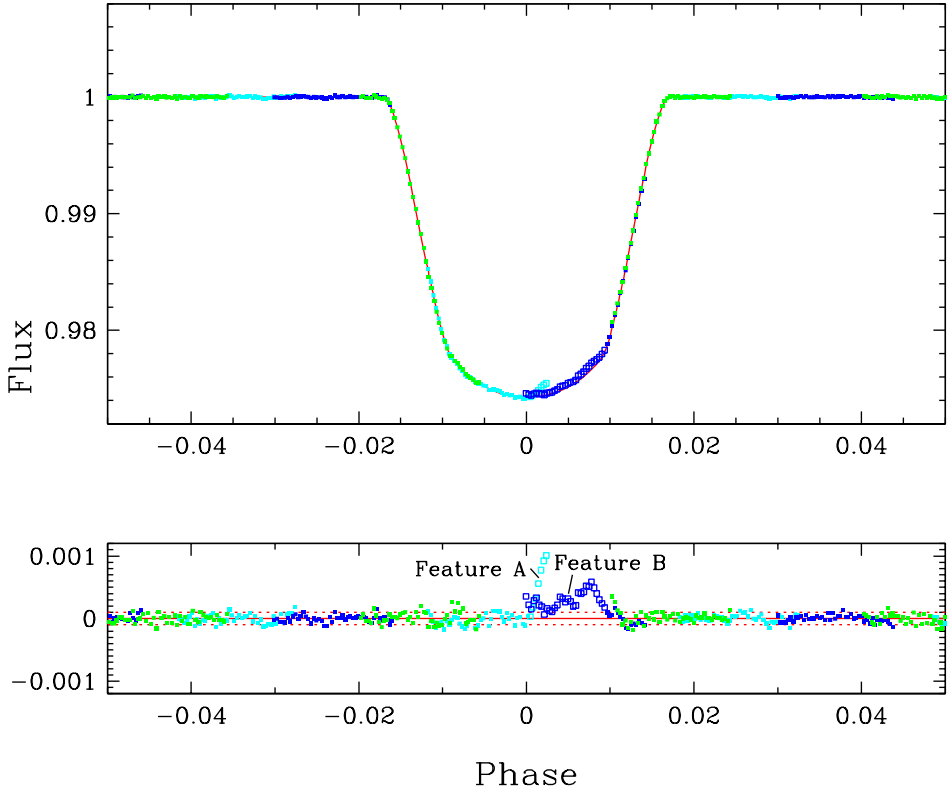}}
\caption{Decorrelated lightcurve phased to P=2.218583 days, with best-fit model transit curve. {\bf 
Top}: flux after external parameter decorrelation as a function of phase;
{\bf Bottom}: residuals around the best-fit transit model. Light blue for the first visit, dark blue for the second and green for the third. Open symbols indicate data affected by Features A and B (see text), not used in the fit. Dotted lines show the $\pm 10^{-4}$ level.}
\label{corlc}
\end{figure*}

%
%
%

\section{Results}

\begin{table}
\centering
\begin{tabular}{c c }
Date & Flux \\ \hline
           2453877.71866         &    0.999421  \\
           2453877.71935	 &    0.999318  \\
           2453877.72005	 &    0.999580  \\
           2453877.72074	 &    0.999518 \\
           2453877.72143	 &    0.999498 \\
           2453877.72213	 &    0.999603 \\
           2453877.72282	 &    0.999592  \\
           2453877.72352	 &    0.999637  \\
           2453877.72421	 &    0.999532  \\
           2453877.72491	 &    0.999610 \\
	   . & . \\
 \hline
\end{tabular}
\caption{Photometric time series (electronic table)}
\label{lc}
\end{table}

\subsection{Lightcurve}
 
The lightcurve decorrelated from dependence on instrumental parameters is given 
in Table~\ref{lc}, and shown in Figure~\ref{corlc}. The data in the first 
orbit of each visit, not used in the analysis, are not shown.

The standard deviation of the individual data points outside the transit is $67\times 
10^{-6}$. Expressed in terms of signal-to-noise, this corresponds to 15000:1. 
The photon noise is $45\times 10^{-6}$. The standard deviation on 10-points average is 
$28\times 10^{-6}$, corresponding to a S/N of 35000:1.  

No out-of-transit structure is visible in our lightcurve above the level of
the residual instrumental systematics.

Transit lightcurves are expected to be perfectly symmetrical to better than the 
$10^{-4}$ level. Two asymmetric features are apparent in our data during the 
transit, one at the end of the second orbit of the first visit, the other at the 
beginning of the third orbit of the second visit. We label the first "Feature A" 
in our analysis, and the second "Feature B". 

%
%
%

Feature A is much too large to be explained by an instrumental effect, and no 
special behaviour of the telescope was observed at the corresponding time. The 
flux increase during Feature A is also seen in the zeroth-order image on the 
CCD. It is accompanied by a detectable change of spectral distribution, which 
suggests an explanation in terms of the transiting planet occulting a cool spot on 
the surface of the star (see Section~\ref{spot}).

Feature B is also larger than  instrumental effects, and does not correlate with 
any of our external instrumental parameters. Its less regular shape and the fact 
that any colour effect is below the noise level indicate that, in principle, an 
explanation in terms of instrumental noise cannot be entirely excluded. Based on 
our experience with previous HST high-accuracy times series, and the simulations of Section~\ref{syscor},
we believe, however, that Feature B is also real.

We do not use Features A and B in the analysis of the lightcurve in terms of 
planetary transit. They are treated separately in Section~\ref{spot}.

%
%

\subsection{Stellar activity and variability}

HD 189733 is an active star, variable to the percent level. It is listed in the
Variable Star Catalogue as V452 Vul. A chromospheric activity index of $S=0.525$ has
been measured by Wright et al. (2004). Activity-related X-ray emissivity has been 
measured by both EXOSAT and ROSAT, activity-related radial velocity residuals of 15 $m\,s^{-1}$ were 
reported by Bouchy et al. (2005). Winn et al. (2007) have measured the photometric variability of 
this star extensively and confirm variability at the percent level, 
compatible with an explanation in terms of transient spots modulated by a rotation 
period of $P_{rot}\sim 13.4$ days. Moutou et al. (2007) measure strong activity in the CaII line
and infer a strong magnetic field with a complex topology from spectropolarimetric monitoring.  The explanation of Features A and B in terms 
of starspots is therefore natural. The presence of large starspots is also confirmed
by an observing campaign on this object by the MOST satellite (Croll et al., in prep.). 
The MOST data yield an improved rotation period of 11.8 days.

\begin{figure}
\resizebox{8cm}{!}{\includegraphics[angle=-90]{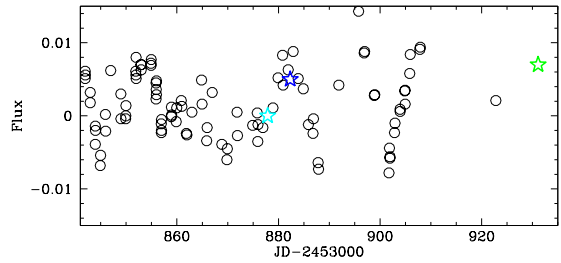}}
\caption{Photometry of HD 189733 from Winn et al. (2007; {\it dots}), with the timing and mean flux level of the three HST visit indicated (shifted by an arbitrary constant; {\it stars}).}
\label{vario}
\end{figure}

The Winn et al. (2007) photometry is contemporaneous with our HST data. Our absolute measurements are placed within the context of the ground-based monitoring in Fig.~\ref{vario}, with an arbitrary zero-point shift. The HST data is in agreement with the periodic variation seen in the long-term lightcurve. If we interpret this variability in terms of starspots moving in and out of view with the rotation of the star, then the third visit occurs near the brightest point -- with less star spots visible -- and the first visit with a 0.007 dimming due to starspots. The phasing and amplitude of features A and B are perfectly compatible with an explanation in terms of the planet occulting part of the starspots responsible for the photometric variation (see Section~\ref{spot}).

Before fitting a transit signal, we correct for the variations of the total stellar luminosity due to the presence of starspots. Outside of features A and B, the planet crosses a spot-free region, therefore a region slightly brighter than the average over the stellar disc, which includes the spots. This is a tiny correction of the scaling between transit depth and radius ratio (of the order of $2\times 10^{-4}$ in flux). Nevertheless, to the level of the accuracy of the HST lightcurve, it makes a significant difference and must be accounted for. We use the absolute flux differences measured in Section~\ref{abs}.

\subsection{Transit signal}

A transit light curve computed with the Mandel \& Agol (2002) algorithm was fitted 
to the light curve, with a downhill simplex algorithm (Press et al. 1992). 
Features A and B were removed with cuts from JD=877.875 to the end of orbit 3 of visit 1, and 
from the beginning of orbit 3 of visit 2 to JD=882.333. The limb-darkening coefficients were left as free parameters, using the four-coefficients non-linear expression of Claret (2000). They converge to $a_{1-4}=(0.506,-0.727,1.345,-0.519)$. These values are compatible with the ones  for $T_{eff}=5000 $K, [Fe/H]=0.0 and $\log g =4.5$ for the R filter.

%

The resulting best-fit transit parameters are shown in Table~\ref{table1}. The best-fit 
theoretical transit curve and the residuals are shown in Fig.~\ref{corlc}. 

\begin{table}
\centering
\begin{tabular}{l  l}
\hline
{\it Measured in this paper} & \\
Period [days] & 2.218581 \ppmm 0.000002\\
Transit epoch [JD] & 2453931.12048 \ppmm 0.00002\\
Radius ratio $R_{pl}/R_*$ & 0.1572 \ppmm 0.0004\\
System scale $a/R_*$ & 8.92 \ppmm 0.09\\
Impact parameter $a \cos i / R_*$ & 0.671 \ppmm 0.008\\ 
Orbital inclination $i$ [$^o$] & 85.68 \ppmm 0.04 \\
Host star $M^{1/3}/R$ ratio & 1.242 \ppmm 0.012\\
Visit 1 transit epoch [JD] & 2453877.87448  \ppmm 0.00005 \\
Visit 2 transit epoch [JD] & 2453882.31171 \ppmm 0.00005 \\
Visit 3 transit epoch [JD] & 2453931.12048 \ppmm 0.00003 \\
 & \\
{\it Inferred} & \\ 
Stellar radius [$R_\odot$]&  0.755 \ppmm 0.011 \\
Planetary radius [$R_J$] & 1.154 \ppmm 0.017 \\
 & \\
\multicolumn{2}{l}{\it Adopted (from Bouchy et al. 2005)}  \\
Stellar mass [$M_\odot$]& 0.825 \ppmm 0.025  \\
Planetary mass [$M_J$]& 1.15 \ppmm 0.04  \\
\hline
\end{tabular}
\caption{Parameters of the transiting system HD 189733.}
\label{table1}
\end{table}

The standard deviation of the residuals around the model lightcurve in the transit are 129$\times 
10^{-6}$. This is larger than the photon noise, and is due to the 
presence of systematic fluctuations on the scale of $10^{-4}$, even outside 
Features A and B. These fluctuations may be due to systematics not perfectly 
corrected by the decorrelation, or to structures on the surface of the star 
smaller in scale than those responsible for Features A and B.

%

Including parts or all of Features A and B in the fit, or changing the limb 
darkening coefficients within reasonable limits, does not modify the resulting 
system parameters beyond the 1\% level, and therefore has no influence on the 
astrophysical applications.

The photon-noise uncertainties on the transit parameters are extremely small, 
and the actual uncertainties are dominated by correlated flux residuals.
 The amplitude of the residuals and the tests in Section~\ref{reduc} 
show that these effects are of the order of $10^{-4}$ at most over the relevant timescales. 
We therefore estimate the uncertainties 
on the transit parameters by allowing the data to move, as a whole, by a maximum 
of $10^{-4}$ in flux from the model, following the approach of Pont, Zucker and Queloz (2006) to 
derive error intervals in the presence of correlated noise. These values can be considered upper limits for 
the uncertainties, since the systematics are unlikely to be exactly in phase 
with the differences caused by changing the transit parameters.

The uncertainties due to these fluctuations, in turn, are dominated in the derivation of the physical 
parameters of the system by the uncertainty on the stellar mass and radius, due 
to the $R M^{-1/3}$ degeneracy of transit systems. As an order-of-magnitude 
indication, the photon noise uncertainty on the radius ratio is $\sim$ 0.01\%,
the uncertainty due to systematics is $\sim$ 0.1 \%,
and the uncertainty due to $R M^{-1/3}$ degeneracy is $\sim$ 1\% (See Section~\ref{trparam}).

%
%
%
%

\subsection{Transit timings}

Table~\ref{table1} also gives the values of the central epoch of the three transits covered by 
our HST data, when fitted independently with the other parameters fixed at the 
values given by the combined data.

\begin{figure}
\resizebox{8cm}{!}{\includegraphics{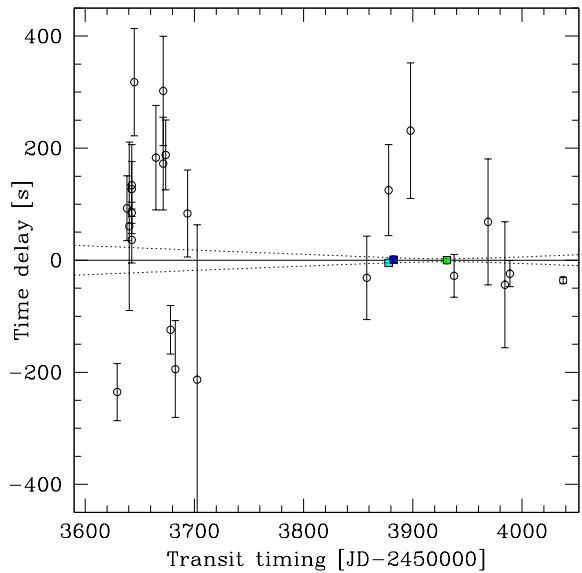}}
\caption{Transit timing residuals for HD 189733, from Bakos et al. (2006), Winn et al. (2007), Knutson et al. (2007) 
and this paper, relative to our best-fit ephemerides. The three HST transits are indicated by the larger squares.The dotted lines delienate our $1-\sigma$ uncertainty interval.}
\label{ttv}
\end{figure}

We compared these values with previous ground-based mesurements (see Figure~\ref{ttv}).
Deviations of transit timings from strict periodicity can reveal the existence of other planets in the system (e.g. Agol et al. 2005, Holman \& Murray 2005). However, the relation between transit timing variations and the orbital elements of the unseen perturber is very complex, and non-detections exclude only restricted regions of parameter space. Periodicity at the level of a few seconds excludes, for instance, the presence of an Earth-mass body in 2:1 resonance with the giant planet (see figure~5 of Agol et al. 2005).
 The HST transit timings are compatible with previous determinations, while 
being much more precise. The periodicity of the transits is stable at the 
level of a few seconds over the three HST transits (the residuals relative to the periodic solution are $5\pm4$ s, $-1\pm 4$ s and $0 \pm 3$ s for the three visits respectively). This strongly suggests that 
the $\chi^2$ excess seen in the transit timing for ground-based data is due to unrecognised 
systematics, as suggested by Bakos et al. (2006), rather than real orbital variations.

The only transit timing of precision comparable to the HST is that of Knutson et al. (2007), the last point on Figure~\ref{ttv}. It is incompatible with the HST timings at the $\sim 4 \sigma$ level. This could be a first indication of transit timing variations, of the order of half a minute. Given the difficulty of estimating systematic errors at this level of accuracy, more data is needed to confirm this tentative indication. An intense monitoring of HD 189733 with the {\it Spitzer} space telescope is currently under way.

%
%
%

%

\subsection{Moons, rings, and spots}

\begin{figure}
\resizebox{8cm}{!}{\includegraphics{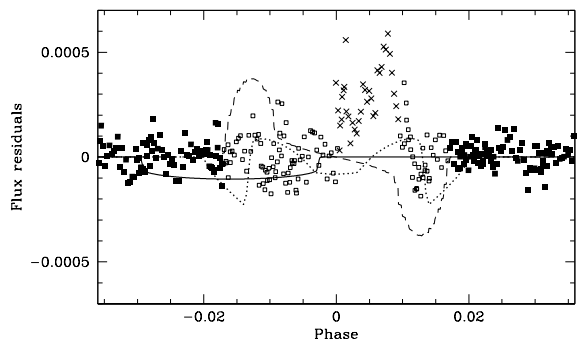}}
\caption{Residuals around the best-fit transit model. Open symbols indicate data during the transit, crosses data attributed to Features A and B. The lines show the signals 
expected from an additional transiting 0.8 Earth-radius body (solid), from planetary rings (dotted), 
and from transit timing variations of 20 seconds (dashed). }
\label{residuals}
\end{figure}

No evidence is seen in the flux residuals for the characteristic signature of: (a) a transiting satellite or 
second planet (b) planetary rings (c) planet or star oblateness (d) transit 
timing variations. The residuals around the best-fit transit lightcurve are compared to the expected shapes for three effects in Fig.~\ref{residuals} (the effect of planet oblateness resembles that of rings).

The two asymmetrical features observed during the transits do not have the 
correct shape to be explained in terms of the transit of a second body in the 
system. They do not occur at the phase expected for such effects as the presence 
of rings, non-sphericity of the star or planet. Feature~A and B are best 
explained as due to the passage of the planet in front of cooler regions -- 
starspots or spot complexes -- on the surface of the star itself.

Feature A, to first order a linear rise in flux, can be well explained by the 
planet occulting part of a large cooler region on the star. Feature B is less regular and has 
lower S/N. Its exact shape is sensitive to uncertainties on the limb darkening parameters and the absolute flux difference between the HST visits.

\begin{figure}
\resizebox{8cm}{!}{\includegraphics{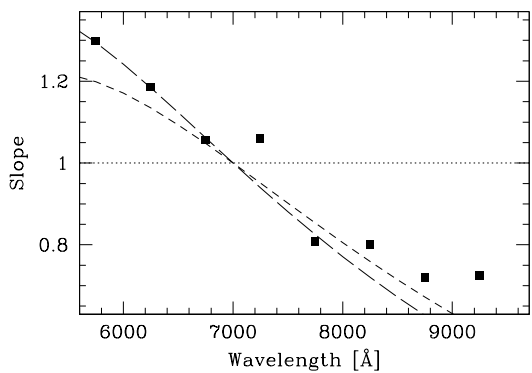}}
\caption{
Slope of the residual flux rise caused by ``Feature A'' in the time series of 500 \AA\ spectral bins, divided by the slope of the lightcurve integrated over all wavelengths. The lines show the expectation for a spot 1000 K cooler than the surface of the star (short dashes), 100 K cooler (long dashes), and for systematic noise without a colour signature (dots).  
}
\label{spotA}
\end{figure}

\label{spot}


For the duration of Feature A, the planet moves $\sim 80\, 000$ km across the face of 
the star. The orbital and rotation axes of the system are almost aligned (Winn 
et al. 2006) and the rotation velocity of the star is long compared to the 
duration of the transit (Winn et al. 2007), so that this distance corresponds to 
a similar projected distance in longitude on the surface of the star. This gives the minimal 
extension in longitude of the spot responsible for Feature A. 

The minimum latitude extention of the spot is given by the total flux increase 
of Feature A. If the spot is much cooler than the rest of the star surface, the 
flux it emits is negligible, and it can be considered a dark feature. In that 
case, the width needed to produce the observed flux increase is $12\, 000$ km. 
 At the other extreme, the occulted portion of the 
spot cannot be larger than the diameter of the planet, R$\sim \! 165\, 000$ km. To 
produce the observed drop in flux, a spot of this size will need to have a 
temperature less than 100~K  degrees cooler than the rest of the star.

There is one additional constraint in the data to support the spot interpretation. Since a starspot is cooler than the surrounding regions on the star's surface, the occultation of a spot will introduce a wavelength dependence in the flux rise. The flux rise will be steeper in the blue, where the temperature dependance is sharper, than in the red. The magnitude of this effect will depend on the effective temperature of the spot compared to the rest of the star. 

We computed the flux in our spectra in 500-\AA \ wide spectra bins, from 5500 to 9500 \AA\ (decorrelated for systematics with the procedure explained in Sect.~\ref{reduc}). For each bin, we fitted a transit model lightcurve with the parameters fixed as per Table~\ref{table1}, leaving only the limb-darkening coefficients free. We then fitted a linear slope to the flux rise caused by Feature A. We divided this slope by the slope measured on the integrated lightcurve. Figure~\ref{spotA} shows the value of this slope as a function of the central wavelength of the spectral bin. We compare these to the expectations for a signal caused by occulting a cooler region (dashed lines in the figure), using the blackbody approximation for the spectral distribution of flux in the stellar atmosphere inside and outside the spot and setting T=5000~K for the stellar surface outside the spot. The data is clearly compatible with the spot scenario, and a colour-independent systematics (dotted line in the figure) can be excluded. 

In principle, this procedure could also allow us to measure the temperature of the spot. But in this range of temperatures and spectral coverage, a bigger, warmer spot and a smaller, cooler spot produce very similar signatures, that are within the observational uncertainties.

\begin{figure*}[thp!]
\resizebox{16cm}{!}{\includegraphics{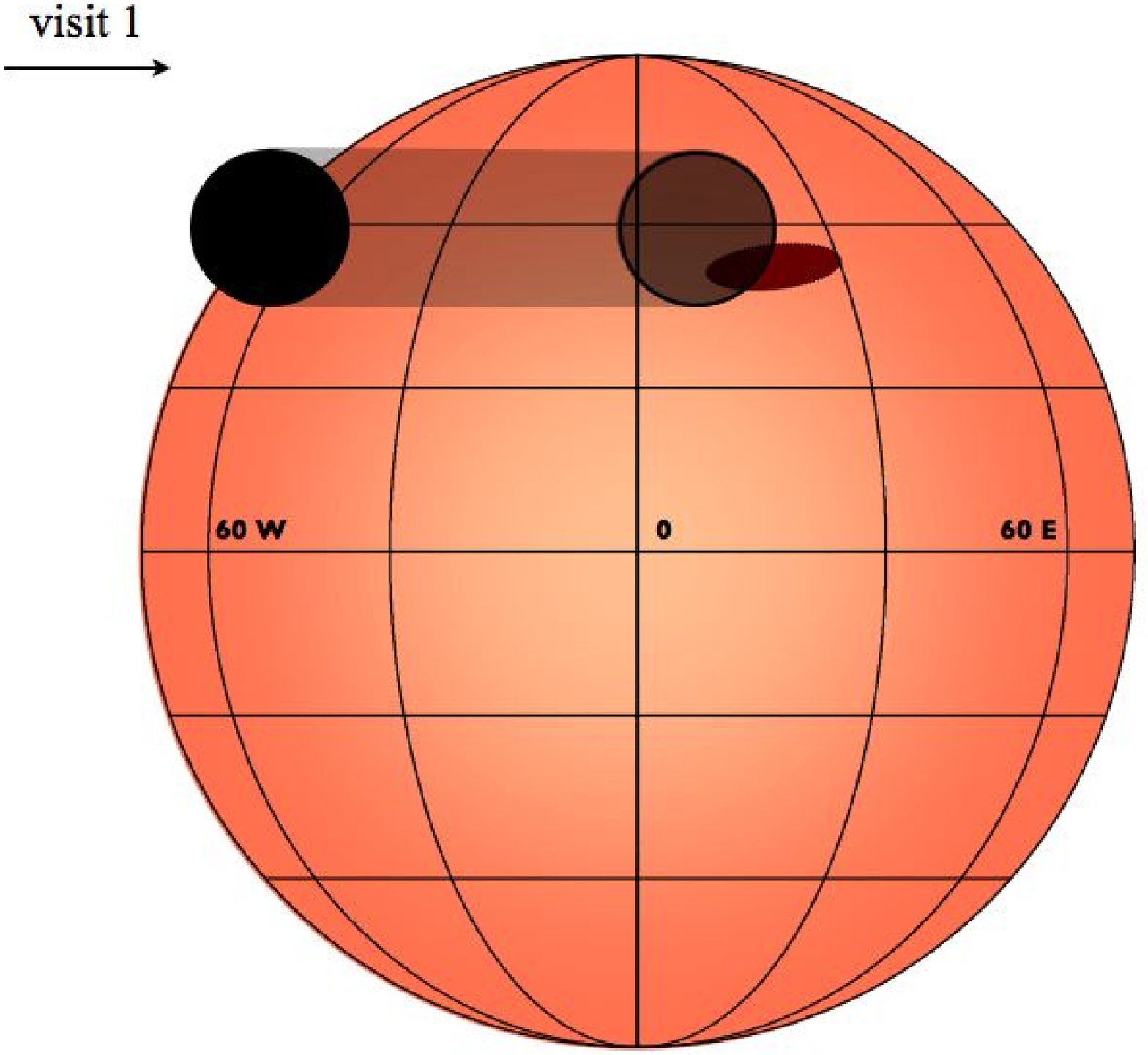}\includegraphics{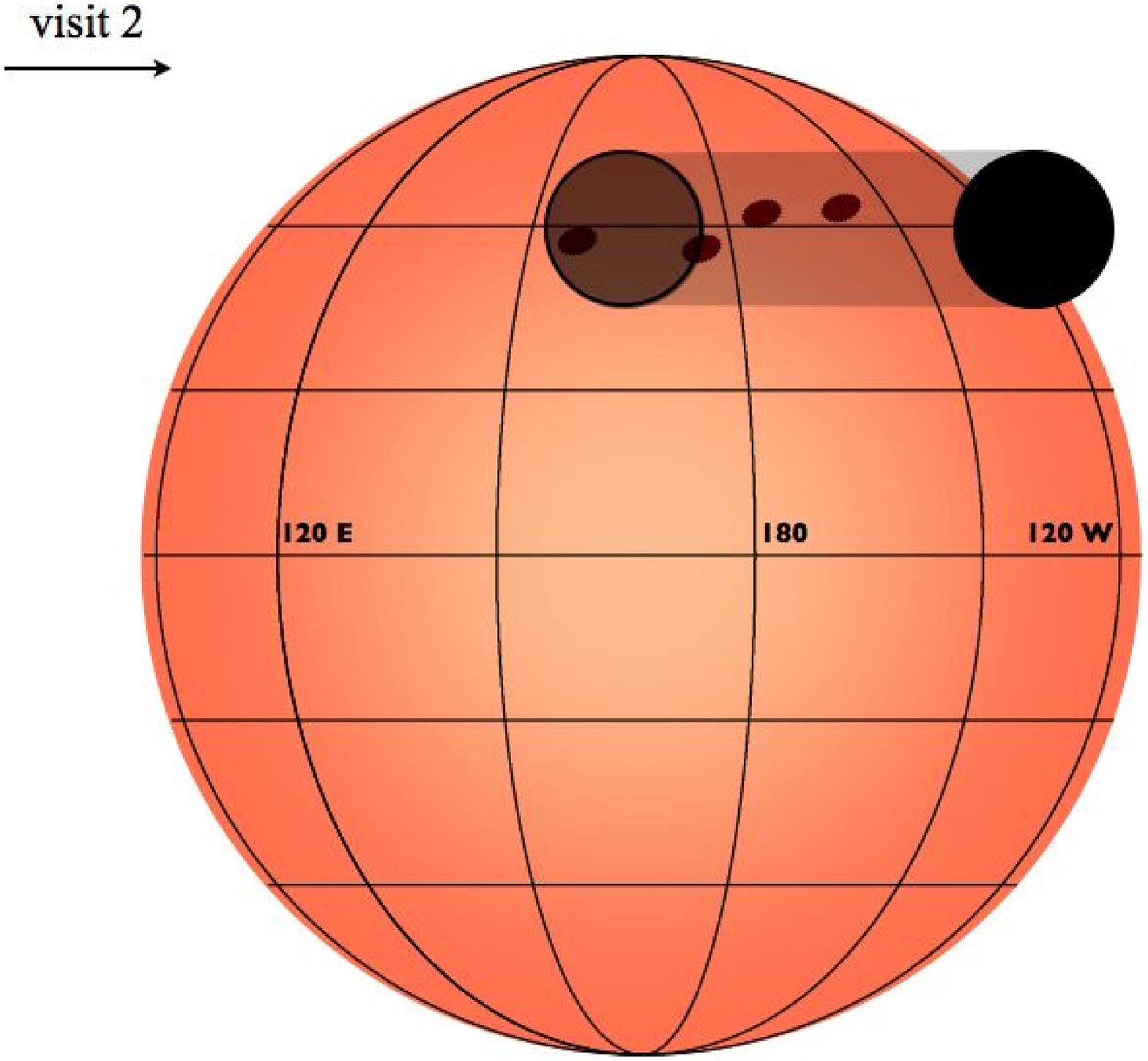}\includegraphics{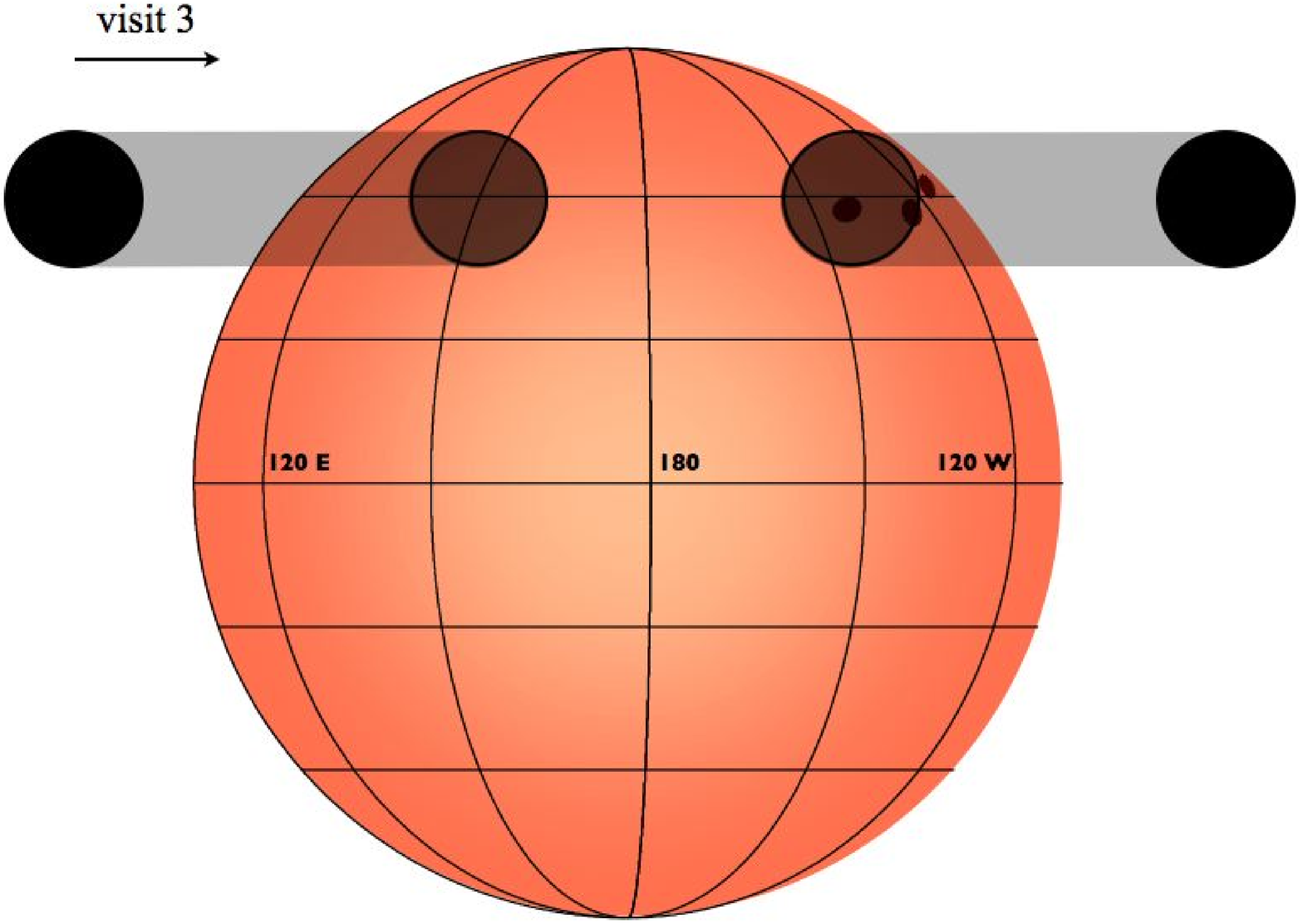}}
\caption{Schematic configuration of the position of the star and planet during the three HST visits. Longitude and latitude lines are drawn on the star at 30 degrees intervals. Longitude zero is defined as the longitude pointing towards the Earth at the epoch of the first transit. The arrows indicate the direction of the stellar rotation and planet orbital motion. The path of the planet during the HST observations is shown for the three visits. Darker features on the star sketch the configuration for the spots affecting the HST lightcurve.}
\label{starmap}
\end{figure*}

In summary, Feature A is compatible with a complex of cool spots extending at 
least $80\, 000$ km in longitude and $12\, 000$ km is latitude, in a relatively uniform 
manner, with an overall effective temperature much lower than the 
rest of the star. It is also compatible with a larger, more circular spot, with 
a smaller temperature difference, down to about $\Delta T < \sim $ 100 K for a spot as 
wide or wider than the planet.

The same procedure was repeated for Feature B. However, in this case, the total 
amplitude of the effect is too small, and any colour signature is dominated by 
the uncertainties. Also, the more complex shape of the signal is compatible with 
several arrangements of spots. Therefore, while we posit that the flux variations 
corresponding to Feature B are due to the occultations of activity-related 
features on the surface of the star, we do not attempt to model this feature in detail.
 In the limit of large temperature difference (dark spots), 
the size of the spots responsible for Feature B would have to be equivalent to a circle of $9\, 000$ km
radius (more than one spot would be necessary to reproduce the observed lightcurve shape).

Figure~\ref{starmap} illustrates the geometric configuration of the star and planet as seen from Earth during the three HST visits, giving a schematic indication of the position and size of the spots. The path of the planet is shown only during HST observations (at other times the HST line of sight is blocked by the Earth). A rotation period of 11.8 days is used to rotate the star between the different visits. The rotation of the star during the transits is very small (about 2 degrees), and can be neglected. There is a global north-south degeneracy. The longitudinal extension of the spots is well constrained, but not their exact sizes and latitudinal position relative to the planet. The stellar rotation is aligned with the planetary orbit, as measured by Winn et al. (2007) using the Rossiter-McLaughlin effect. 

The spot responsible for Feature A is not encountered again during the two other visits. However, the spot complex causing Feature B is transited again during egress in visit 3. This may explain the behaviour of the residuals at that point in the lightcurve (see bottom panel of Fig.~\ref{corlc}). Altogether, the planet transits about 10\% of the star's surface during our observations, and occults starspots about half this time, suggesting that spots of different sizes are abondant on the stellar surface.

\subsection{Host star and planet}

%
\label{trparam}

The radius of the primary and its mass to the one-third power are degenerate in 
transit lightcurves.  Second-order effects that 
would allow breaking this degeneracy, such as the light travel-time effect, are far too small to 
be detectable in the presence of any realistic amount of photometric noise or 
stellar variability. Therefore, the $M^{-1/3}\, R$ degeneracy must be lifted 
with recourse to stellar evolution models.

\begin{figure*}
\resizebox{16cm}{!}{\includegraphics{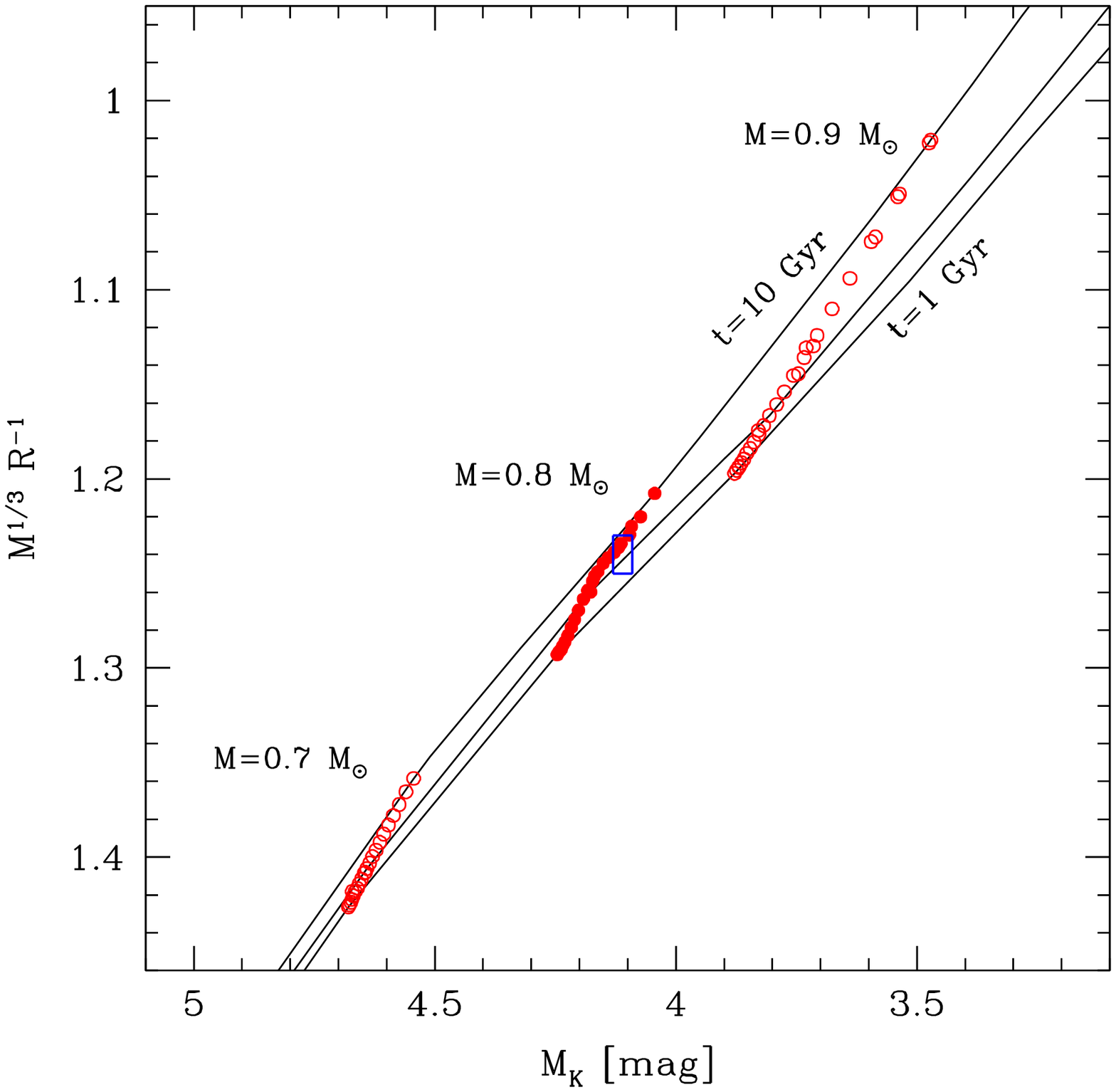}\includegraphics{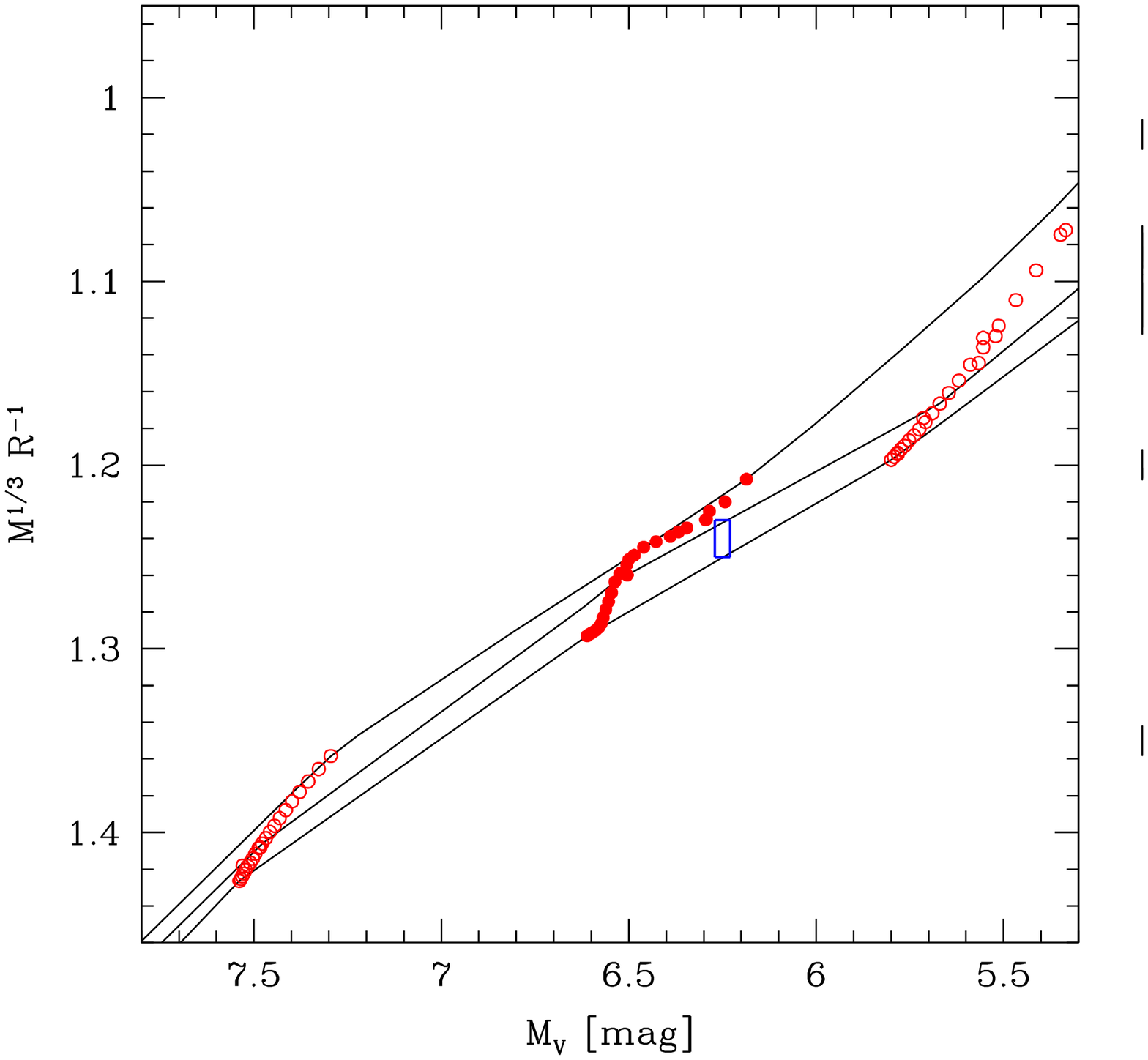}}
\caption{Constraints on HD 189733 from the absolute visible and infrared magnitude, and the $M^{1/3}\,R^{-1}$ ratio, compared to Padova stellar evolution model isochrones.}
\label{padova}
\end{figure*}

Fortunately, HD 189733 is a well-observed, low-mass star, and evolution 
models allow for only a small range of masses compatible with its observed 
visible and infrared magnitude, parallax, and spectroscopic temperature. Its 
Hipparcos parallax of 51.94 \ppmm 0.87 $mas$ implies a distance of 19.25 \ppmm 0.32 parsec, 
and absolute magnitudes of 
$M_V$=6.25 in the visible and $M_K$=4.07 in the infrared. These luminosities can be 
combined with the $M^{-1/3}\, R = 1.246 \pm 0.012$ (in Solar units) value
 from the HST transit shape and 
confronted to stellar evolution models. Figure~\ref{padova} shows the error box for 
HD189733 in these two projections of parameter space compared to the position of Padova stellar 
evolution models (Girardi et al. 2002). The data are coherent with the models and 
compatible with a mass in the 0.80-0.85 $M_\odot$ range for HD 189733. 
Similar results are obtained using the Baraffe et al. (1998) low-mass star models. We therefore 
confirm the value of $M=0.825 \pm 0.025 \ M_\odot$ used in Bouchy et al. (2005) for the 
mass of the host star.

%

It is interesting to note that low-mass eclipsing binaries with well-determined 
mass and radius are not compatible with the model tracks in this range of mass (Ribas et al. 2006), 
unlike HD 189733. This is a strong indication confirming the suspicion that 
close eclipsing binaries strongly influence each other, and are therefore not 
good calibrators for the mass-radius relation of single stars.


We adopt $M=0.825\pm 0.025 M_\odot$ to lift the mass-radius degeneracy, which implies 
$R_{star}=0.753 \pm 0.011 R_\odot$, and $R_{pl}= 1.147 \pm 0.017 R_J$ (with $R_J = 71500$ km). These values are listed in Table~\ref{table1}.

The planetary mass remains unchanged at $M_{pl} = 1.15 \pm 0.04 $ from Bouchy et al. (2005).
%
%
%
%
This corresponds to a planetary density of 822 kg m$^{-3}$. 
%
%

%
%

%
%

%

\section{Discussion}

The ACS spectrophotometric time series presented in this article shows the 
capacity of HST to obtain lightcurves of extreme accuracy for bright stars. It echoes
the HST lightcurve of Brown et al. (2001) acquired with the STIS spectrograph for 
HD 209458. The S/N for HD 189733 is even higher because of the larger wavelength domain
covered (at lower spectral resolution), the fact that HD 189733 is slightly brighter 
in the wavelength interval sampled, and the higher throughput of ACS/HRC.  

The single-point standard deviation outside the transit is $67 \times 10^{-6}$, 
for a sampling of one 
point per minute. Correlated systematics left after
decorrelation of instrumental parameters such as tracking drift and focus change
have an amplitude well below the $10^{-4}$ level. The standard deviation of the
mean of measurements over 10 minutes is $28 \times 10^{-6}$. By comparison, systematics in 
high-precision ground-based photometry are typically of the order of one part per 
thousand, or $1000\times 10^{-6}$.

%

%

These data allow a very precise measurement of the planetary radius to be 
obtained. This is also due to the fact that the constraints on the mass of the 
star are good, because it is a low-mass star with a known parallax from 
Hipparcos.
%


The HST/ACS lightcurve permits to lift the degeneracy between primary radius and 
orbital inclination in the transit signal. Bakos et al. (2006) show how the data of a 
ground-based photometric campaign, even with several telescopes and in several 
filters, can lead to misleading conclusions in this respect. The Bakos et al. 
data apparently indicated a much smaller primary ($R_*=0.68 \pm 0.02\ R_\odot$), 
but the authors concluded that 
this was due to unrecognized systematics in the ground-based photometry. We reinforce 
this conclusion. Our results, on the other hand, confirm the values of 
Winn et al.(2007), with smaller uncertainties. We conclude that Winn et al. shows the 
proper amount of data necessary to lift the radius-inclination degeneracy from 
the ground with confidence, for such a deep transit: 8 complete transits covered in 
excellent conditions. This gives a yardstick 
against which to evaluate similar ground-based campaigns.

More recently, Baines et al. (2007) have measured the radius of HD 189733 directly with the CHARA interferometer, obtaining $R_*=0.779 \pm 0.052 R_\odot$. This value has higher uncertainties but is completely independent of any assumption of mass, and is therefore a useful check to the determinations from the geometry of the transit.
It is compatible with our value.

From the 8-$\mu$ transit lightcurve measured with the {\it Spitzer} space telescope, Knutson et al. (2007) obtain $R_{\rm pl}/R_{\rm star}= 0.1545 \pm 0.0002$. This is significantly lower than our value. Starspots not crossed by the planet during the transits can contribute to this difference. These spots will lower the surface brightness of the part of the star not occulted by the planet, therefore increasing the flux drop and the apparent radius ratio. This effect is smaller in the infrared, because the spot contrast gets smaller. Another possible explanation is that the effective transit radius of the planet is indeed larger in the visible and near infrared than in the mid-infrared, due to the transmission spectrum of the planetary atmosphere. These topics are beyond the scope of this paper.

We here quantify our non-detection of moons or rings in terms of upper limits to 
any satellite or ring system around HD 189733b. 

We have an upper limit of $10^{-4}$ to the depth of any unrecognised second 
eclipse in our data. This corresponds, given the size of the primary star, to a 
body of 5000 km (0.8 Earth radii). Unless the satellite spends each of the three 
transits that we have sampled either in front or behind the planet -- an 
extremely unlikely configuration - we conclude that a moon larger than 0.8 
$R_\oplus$ around HD 189733b is excluded by the data. Figure~\ref{residuals} shows the signal such a
moon would cause, superimposed to the residuals during the three HST visits.

We also exclude a ring system that would cause a flux drop larger than $10^{-4}$ 
around the time of transit. Since the planet is orbiting very close to the star 
and is subject to strong tides, we assume that any ring system would be aligned 
with the orbital plane ($i= 85.72 $). In that case, a Saturn-type ring system at 
about 1 $R_{pl}$ from the star's surface will need to be less that  $120\, \tau$ km 
wide, where $\tau$ is its optical depth (if Saturn's 
ring are any guide, a ring system should be thin enough that the ring's 
thickness be negligible compared to its projected width even for the low inclination 
of HD 189733, so that this result is valid for any viewing angle of the rings except exactly edge-on).  
For comparison, the Saturn main ring system is $40\, 000$ km wide with 
a high optical depth ($\sim 0.3$), and the Jupiter ring system is also thousands of km wide but has a very small optical depth ($10^{-6}$ to $10^{-7}$). We can 
therefore exclude a major debris ring around HD 189733b (discounting the unlikely coincidence of an exact alignement with the line-of-sight).

We note that altogether these limits would not have allowed us to detect any of 
Jupiter's moons, and Jupiter's weak ring system. Therefore, the absence of moon 
and rings in HD 189733 shown by our data must be understood as showing the 
absence of major, Earth-size moons and Saturn-type debris rings.

%

The fact that spot-like features affected a large part of the HST transits suggests that activity-related features are abundant on the surface of HD189733.
The three HST visits together occulted slightly less than 10\% of the star's surface, revealing one spot at the $10^{-3}$ level and one at the $4\cdot 10^{-4}$ level. This is in good order-of-magnitude agreement with the variability observed by Winn et al. (2007), which changes from negligible to 1\% over a few rotation cycles. 

A coherent picture of the active surface of HD189733 emerges from the combination of the small-scale information from HST data and the large-scale variability monitored by Winn et al. (2007). The timing of the sudden flux rise seen at the end of the first HST visit (``Feature A'') is compatible with the planet occulting part of a large spot responsible for a significant fraction of the modulation of the lightcurve. Between the first and the second visit, the luminosity of the star has increased by 0.5\%, and about a third of the stellar rotation period has elapsed, giving the spot (or spots) enough time to rotate from slightly east of the center of the star to the hidden side.

In conclusion, our results demonstrate again the potential of HST to gather extremely precise lightcurves for bright stars. We derive improved new measurements of the characteristics of this important system, that provide very tight constrains for the models. Our lightcurve did not show any evidence for Earth-like satellites or rings around the planet HD 189733, nor transit timing variations indicative of an unseen second planet. The lightcurve during the transit was strongly affected by the spots and active regions crossed by the planet, providing a rare glimpse of the small-scale geometry of spots on a star other than the Sun.

\begin{acknowledgements} 

We thank Eliza Miller-Ricci and Heather Knutson for very helpful discussions. We are indebted to the StScI staff for their remarkable handling of the scheduling of our time-critical observations. 

\end{acknowledgements}

\section*{Bibliography}

Agol, E.; Steffen, J.; Sari, R.; Clarkson, W. 2005, MNRAS 359, 567
\\ \noindent
Baines, E. K. ; van Belle, G. T. ; ten Brummelaar, T. A. ; McAlister, H. A. ; Swain, M.; Turner, N. H. ; Sturmann, L. ; Sturmann J 2007, ApJ Letters 661, 195
\\ \noindent
Bakos, G. ç.; Knutson, H.; Pont, F.; Moutou, C.; Charbonneau, D.; Shporer, A.; Bouchy, F.; Everett, M et al. 2006, ApJ 650, 1160
\\ \noindent
Baraffe, I.; Chabrier, G.; Allard, F.; Hauschildt, P. H. 1998, A\& A 337, 403
\\ \noindent
Brown, T. M.; Charbonneau, D.; Gilliland, R. L.; Noyes, R. W.; Burrows, A. 2001, ApJ 552, 699
\\ \noindent
Claret, A. 2000, A\& A 363, 1081
\\ \noindent
Deming, D.; Harrington, J.; Seager, S.; Richardson, L. J. 2006, ApJ 644, 560
\\ \noindent
Girardi, L.; Bertelli, G.; Bressan, A.; Chiosi, C.; Groenewegen, M. A. T.; Marigo, P.; Salasnich, B.; Weiss, A. 2002, A\& A  391, 195
\\ \noindent
Grillmair, C. J.; Charbonneau, D.; Burrows, A.; Armus, L.; Stauffer, J.; Meadows, V.; Van Cleve, J.; Levine, D. 2007, ApJ 658, 115
\\ \noindent
Holman, M.J.; Murray N.W. 2005, Science 307, 1288
\\ \noindent
Knutson, H.A.; Charbonneau, D.; Allen, L.E.; Fortney, J.J.; Agol, E. et al. 2007, Nature 447, 183
\\ \noindent
Mandel, K.; Agol, E. 2002, ApJ 580, 171
\\ \noindent
Moutou, C.; Donati, J.-F.; Savalle R. et al. 2007, A\&A in press
\\ \noindent
Pont, F.; Zucker, S.; Queloz, D. 2006, MNRAS 373, 231
\\ \noindent
Press, W.H.; Teukolsky, S.A.; Vetterling, W.T. 1992, ``Numerical recipes in FORTRAN'', Cambridge University Press
\\ \noindent
Ribas, I. 2006, Ap \& SS 304, 89
\\ \noindent
Winn, J. N.; Johnson, J. A.; Marcy, G. W.; Butler, R. P.; Vogt, S. S.; Henry, G. W.; Roussanova, A.et al. 2006, ApJ 653, 69
\\ \noindent
Winn, J. N.; Holman, M. J.; Henry, G. W.; Roussanova, A.; Enya, K.; Yoshii, Y.; Shporer, A.; Mazeh, T. et al. 2007, AJ 133, 1828
\\ \noindent
Wright, J. T.; Marcy, G. W.; Butler, R. Paul; Vogt, S. S. 2004, ApJS 152, 261
\end{document}